# INTERPRETATION OF CONFOCAL ISO 21073: 2019 CONFOCAL MICROSCOPES – OPTICAL DATA OF FLUORESCENCE CONFOCAL MICROSCOPES FOR BIOLOGICAL IMAGING - RECOMMENDED METHODOLOGY FOR QUALITY CONTROL

**Glyn Nelson[1], Laurent Gelman[2], Orestis Faklaris[3], Roland Nitschke[4], Alex Laude[1]**

1. The BioImaging Unit, William Leech Building, Newcastle University, Medical School, Framlington Place, Newcastle Upon Tyne, NE2 4HH, UK.
2. Friedrich Miescher Institute for Biomedical Research, Maulbeerstrasse 66, CH-4058, Basel, Switzerland.
3. MRI-CRBM, CNRS, 1919 route de Mende, 34293, Montpellier, France.
4. Life Imaging Center, Albert-Ludwigs University Freiburg, Habsburgerstr. 49, 79104 Freiburg, Germany.

## NOTES & DISCLAIMER

- In the confocal ISO 21073 (https://www.iso.org/standard/69820.html), sampling rate is very high. The ISO requires 10x oversampling for the PSFs rather than the more usual 3x Nyquist sampling criterion usually employed.
- The first 4 QC metrics listed here satisfy the confocal ISO tests. There are also two other system metrics that should be recorded to fulfil the confocal ISO requirements they are: *recording of the scan optics field number* and the *scanning frequency*. Since both of these values are recorded in image metadata already for many point-scanning systems, they do not need to be manually determined. However, if you have a home-built system, you will need to calculate and record these values, this is described in the Confocal ISO 21073 document.
- Two pieces of software work well for analysis of the above metrics: MetroloJ and PSFj (https://imagejdocu.tudor.lu/plugin/analysis/metroloj/start and http://www.knoplab.de/psfj/overview/ respectively). These are suggested as a suitable method for the analysis of each QC test. It should be noted that the manuals of both pieces of software provide further background and describe the tests in greater detail: they also provide methods for sample preparation and recommend additional literature.
- Where we feel that insufficient information or explanation has been given in the Confocal ISO 20173 document, we have provided supplementary information and have made practical suggestions.
- The measurements listed below we suggest should be performed in order of priority and have the potential to have the greatest impact on system performance. The first four are mandatory requirements for the confocal ISO (ISO 21073) and are our summaries and interpretations of what is described in the confocal ISO 21073 document.
- **Disclaimer:** It should be noted that this document is our interpretation of the Confocal ISO 21073 document and should be taken together with it when developing your own confocal QC methods.

## INTRODUCTION

The performance of a confocal imaging system may be no better than a general-purpose widefield system if it is not properly maintained or quality controlled. Quantitative fluorescence microscopy techniques necessitate the standardisation of preparatory and acquisition protocols but, understanding the performance of the microscope itself, arguably as important for such measurements, has hitherto received little attention.

Maintenance of advanced fluorescent microscopes is essential to allow researchers to have full confidence in the imaging data collected. Standardisation for fluorescence microscopy is poorly practised in the scientific community (1). More emphasis is now being placed on quantitative imaging approaches to answer biological questions (2). The quality of these observations and data are only as good as the quality of microscope used to make them: thus, it is important that microscope 'quality' is understood and documented to support quantified imaging data.

The International Organisation for Standardization (ISO, https://www.iso.org/) has recently created standards for confocal microscopy (3). These ISO standards direct researchers about what should be measured and tested. Nevertheless, within these documents, there is little detailed information describing how key measurements should be made, using which samples/ tools and how often to make them.

To overcome the above deficiencies in microscope Quality Control (QC) and encourage good practise for the publication of quantified microscopy data, we here present methodology to monitor the most important factors influencing the Quality of confocal images. The methods presented also fulfil the requirements of the Confocal ISO. When drawing up this document in response to the Confocal ISO, it became apparent to us that further work was required to determine best methodology and also further minimal QC tests. It was also apparent that these Methods needed to be extended to cover other common imaging modalities. As a consequence, we have now established a multinational, open group of interested parties to take this work further: Quality Assessment and Reproducibility for Instruments and Images in Light Microscopy, (QUAREP-LiMi) initiative (https://quarep.org). It is intended that this group will produce a more definitive and expansive QC Methodology manual for light microscopy to supersede this document.

## ESSENTIALS FOR OBTAINING RELIABLE QC DATA:

1. Ensure you use the correct immersion medium.

2. Ensure the correct cover glass type and thickness (No. 1.5H (170 µm) for most objectives).

3. Ensure good temperature stability in the room and around the microscope

4. Switch the whole system on for at least 60 minutes prior to use to allow lasers and any coolants to stabilise and the detectors to reach the optimal temperature.

# QC MEASUREMENTS

## 1. STABILITY OF ILLUMINATION POWER (CONFOCAL ISO21073, SECTION 4.4)

Comparison of fluorescence intensity between images requires knowledge that the excitation light sources are stable. Ideally this should be recorded to ascertain impact on both short-term (2D) and long-term (3D and time-lapse) imaging (Confocal ISO 21073 Section 4.4) and should be measured using a calibrated external power sensor.

*[Note: We have found that slide-based sensor formats are the easiest to mount (e.g. Thorlabs S170C or Exelitas XP750).]*

### METHOD

1. Ensure lasers have been warmed up prior to measurement for at least one hour.

2. A suitable power sensor is placed in the focal plane of the standard 10x objective on the system (usually air, 0.3-0.4 NA). Ensure background light is kept to a minimum (room lights off).

3. During the recording period, the laser beam should be stationary and continuously illuminate the sensor (i.e. no blanking) but not saturate it.

4. Determine a suitable laser power output (usually by % AOTF) that lies between 0.2 to 0.5 mW at the power sensor.

5. For short-term stability testing, set the scan mode to spot (stationary) scanning and record **every second for 5 minutes**. Calculate stability (STAB$_{short}$) as:

    a. STAB$_{short}$ (%) = 100x (1 - ((P$_{max}$ - P$_{min}$) / (P$_{max}$ + P$_{min}$)))

    b. Where max and min are the maxima and minima recorded during the 5 minutes recording.

6. For long-term stability, spot scanning is performed **every 30 seconds for 1 second over 120 minutes**, recording the power at each imaging timepoint. Calculate stability (STAB$_{long}$) as:

    a. STAB$_{long}$ (%) = 100x (1 - ((P$_{max}$ - P$_{min}$) / (P$_{max}$ + P$_{min}$)))

    b. Where max and min are the maxima and minima recorded during the 5 minutes recording.

*[NOTE: If the software allows, the microscope can be set up to provide alternate laser values over the same time period, scanning each laser sequentially within the 30 second time lapse.]*

7. These data should be reported along with the laser wavelengths measured and the average laser power measured.

## 2. UNIFORMITY OF FIELD CENTRING ACCURACY (SYSTEM FLATNESS OF FIELD) (CONFOCAL ISO21073, SECTION 4.2)

### OVERVIEW

This allows monitoring of objective quality and internal light path optics (e.g. galvo mirrors or filters). Comparison of field flatness over time for each objective will identify any new aberrations in the system (e.g. damaged objective, light source centricity (widefield), offset galvo mirrors). The data can be used for post-acquisition correction if required, and therefore improves the quantification of data from across the field of view, improving data reliability.

### METHOD

1. Prepare a sample consisting of a homogenous fluorescent specimen- this is simply achieved using a solution of dye contained with the correct coverslip for your objective on one side to image through. This is relatively easy to prepare for inverted microscopes (using a glass bottomed dish or well plate), but more awkward for uprights. In this instance, we recommend using a small well made from either a wax circle, a silicone gasket or a geneframe (ThermoFisher Cat No. AB0576) on a slide with a coverslip mounted on top and then sealing with nail varnish.

*[Note: this is the recommendation from Confocal ISO- for point scanning confocals and widefields we have routinely used fluorescent plastic slides instead].*

The choice of dye used is not very important, if it bleaches quickly, there should be enough diffusion of 'fresh unbleached fluorophore' to replace quenched molecules. FITC is cheap and easily obtainable (e.g. search for fluorescein leak tracing dye and several sellers offer it for a few £s or $s).

2. Ensure that the pinhole is the same as for the lateral and axial resolution QC image acquisition (usually 1 Airy Unit) and a minimal scan resolution of 128x128 pixels is set. Ensure the objective is focussed within the homogenous fluorescence of the specimen. Set the excitation and detector to ensure a good signal to noise without saturation.

3. Scan a single plane through the sample in the centre of the field. This data can be analysed using the ImageJ/ Fiji plugin MetroloJ and the field illumination report function. Further details and the manual are available at: https://imagejdocu.tudor.lu/plugin/analysis/metroloj/start. This will provide values for the uniformity of field and the centring accuracy. Recorded metadata should describe:

    a. Field scan optic field number (i.e. objective magnification and scan zoom)

    b. Excitation wavelength and emission wavelength band

    c. Objective details (power, NA, plan, apo etc.)

    d. Pinhole size in Airy nits, plus the wavelength used for calculating this.

## ALTERNATIVE METHOD SUGGESTION

Using an Argolight slide (http://argolight.com/) or PSFcheck slide (https://www.psfcheck.com/), the field of rings provides a metric for measuring flatness of field. Capturing this with all available excitation options allows to check for chromatic aberration across the field of view at the same time too. This should be performed for all wavelengths and all high-end objectives on the system.

## 3. SYSTEM CHROMATIC ABERRATION AND CO-REGISTRATION ACCURACY (CONFOCAL ISO21073, SECTION 4.3)

## OVERVIEW

Co-registration is described as the precision with which a microscope images a fluorescent object with different excitation and emission wavelengths in the same position within the image. It can also be described as the difference between the lateral intensity maxima of multi-spectral images captured of the object imaged in the ***centre of the field***. The multi-spectral object illuminated should reflect in its excitation properties the spread of excitation wavelengths achievable with the microscope system. Co-registration differences mainly arise from chromatic aberration within the objective but can also arise from misaligned optics elsewhere in the system. Excitation should be using the internal light sources (lasers) in combination with a fluorescent sub-resolution multi-spectral bead sample or an etched fluorescent pattern. This should be placed in the objective focal plane.

## METHOD

1. Prepare a suitable sample. For standard widefield and confocal systems, 200 nm or 100 nm multi-colour beads (TetraSpeck microspheres from ThermoFisher) are suitable (see Appendix: Sample Prep.) and should be dried onto the coverslip (thickness 1.5 (170 µm)) to avoid any mounting medium aberrations. 100 nm bead preparations are ideal for also measuring system resolution, so a preparation of these saves on a separate preparation. To aid focussing (can be difficult with such small beads), a more dilute volume of larger (e.g.5 µm) beads can be added- these are large enough to be obviously the wrong beads for imaging.

2. Find the focal plane of the beads and choose a field with relatively sparse distribution, since the Airy rings will interfere between beads if they are too dense. Ensure you have a bead in the centre of the field.

3. *For acquisition, ensure that pixel sizes are at least **10-fold** smaller than the theoretical lateral resolution (for 1.4 - 1.45 NA oil lens at 520 nm emission this is 160 nm, so pixel size should be < 16 nm). Axial steps should also be 10-fold smaller than the theoretical axial resolution (400 nm for a 1.4 NA oil lens at 520 nm emission). The image intensity should be strong enough to visualise the entire dynamic range down to background noise and up to maximal emission, ensuring no saturation (clipping) of the brightest points. A maximum of 1 Airy unit pinhole should be employed, **ensuring it is calculated at the same wavelength as being imaged.***

4. Multicolour images can be captured either sequentially on the same detector or across multiple detectors either sequentially or simultaneously. Ideally, both methods are employed to ensure correct alignment of the detectors as well as testing the lens apochromaticity.

5. Beads can be analysed by determining the centre of gravity for each spectral channel and the difference in localisation between them. PSFj (http://www.knoplab.de/psfj/overview/) is a suitable piece of software for this analysis. It will automatically find all beads in the field and provide a comparison of the centre of mass for each channel. Additional output from PSFj includes analysis of the co-registration across the Field of View, which, whilst not addressed in the Confocal ISO, is still important for knowing the chromatic aberration of your objective. Alternatively, analysis can be performed using the ImageJ/ Fiji plugin MetroloJ and using the co-alignment report function. Further details and the manual are available at: https://imagejdocu.tudor.lu/plugin/analysis/metroloj/start.

*[NOTE: this test can be performed/generated from the same data as the lateral and axial resolution test below if we ensure that the image capture satisfies the sampling criterion, i.e. use 100nm beads and subsampling as described.]*

6. To ensure satisfaction of the confocal ISO (section 4.3.1), this should be performed only on a bead, or the average of beads, in the centre of the field (so images should be cropped to this for PSFj). Additionally, one must ensure that the following parameters are recorded with the co-registration data:

    a. Information on the excitation wavelengths;

    b. Information on the detection wavelength bands;

    c. Manufacturer's designation of the objective;

    d. Size of detection pinhole in Airy units and reference wavelength used to specify the Airy unit;

    e. Information about detection arrangement, e.g. single detector for all wavelength bands or individual detector for each wavelength band;

    f. Information on whether the intensity centre of gravity of the object or the maximum of the cPSF was used.

## ALTERNATIVE METHOD

An alternative approach to using sub-resolution bead samples can be achieved using an Argolight slide or PSFcheck slide, the 'field of rings' structures etched provide a metric for measuring chromatic aberration / co-registration as the patterns fluoresce across a wide spectrum.

## 4. LATERAL AND AXIAL RESOLUTION (CONFOCAL ISO21073, SECTION 4.1)

### OVERVIEW

This allows monitoring of objective quality and other system parameters (e.g. immersion oil). Comparison of PSFs over time for each objective will identify any new aberrations in the system (e.g. damaged/ unclean objective, defective oil). The data can be obtained from the same dataset as the chromatic aberration if multi-wavelength emission diffraction-limited spots / beads are used for that test. The data can be used for post-acquisition correction if required (e.g. deconvolution) and can be reported with published data to prove the resolution and limitations of the system.

## METHOD FOR RESOLUTION ESTIMATION (SECTION 4.1.2)

1. Prepare a bead sample slide with beads mounted as close to the cover glass as possible (see Appendix: Sample Prep) in a mounting medium that matches the lens to be used (e.g. high RI mountant for oil-immersion, glycerol for glycerol-immersion and aqueous medium (in agarose) for water-immersion). Ensure that the mounting medium has cured (if required) for its optimal refractive index (e.g. Mowiol or ThermoFisher Prolong mountants). The beads should be smaller than half the calculated theoretical resolution for the objective / system (e.g. see Appendix 2: Theoretical Resolution)

   *[Note, for high end objectives, e.g. >1.2 NA, this equates to beads smaller than 100 nm. https://www.micromod.de sell green and red fluorescent beads of 10-50 nm diameter, and https://www.cd-bioparticles.com have green fluorescently labelled 50 nm gold nanoparticles- neither tested by us yet. ThermoFisher Fluospheres at 40 nm are also useable and come in a limited range of colours].*

2. Find the focal plane of the beads and choose a field with relatively sparse distribution, since the Airy rings will interfere between beads if they are too dense. Ensure you have a reasonable distribution of beads in the centre of the field. To aid focussing (can be difficult with such small beads), a more dilute volume of larger (e.g. 5 μm) beads can be added- these are large enough to be obviously the wrong beads for imaging.

3. For acquisition, ensure that pixel sizes are at least 10-fold smaller than the theoretical resolution (for 1.4 - 1.45 NA oil lens at 520 nm emission this is 160nm, so pixel size should be < 16 nm), and the field is at least 10-fold larger (1.6 μm for the above example).

*[Note, we recommend capturing a slightly larger field comprising several beads around the centre of the objective to give an average PSF].*

   Axial steps should also be **10-fold** smaller than the theoretical resolution (400 nm for a 1.4 NA oil lens at 520 nm emission). The image intensity should be strong enough to visualise the entire dynamic range down to background noise and up to maximal emission, ensuring no saturation (clipping) of the brightest points. A maximum of 1 Airy unit pinhole should be employed, ensuring it is calculated at the same wavelength as being imaged.

4. For analysis (ISO section 4.1.4), a Gaussian fit should be made for a single pixel width line profile laterally and axially through a bead from the centre of the field. This can be performed in ImageJ from line profile data (Analyze-Tools-Curve Fitting).

5. To satisfy the confocal ISO (section 4.1), confocal FWHM should be reported alongside other metadata:

   a. excitation wavelength,

   b. detection wavelength band,

   c. objective designation from manufacturer,

d. pinhole size in Airy units (plus the reference wavelength used to calculate its size),

e. Polarisation state of excitation light plus its direction relative to the axis of the determined lateral FWHM [NOTE: For high NA objectives and linear polarized excitation light, the FWHM in the direction perpendicular to the direction of the polarization is smaller than the FWHM in the direction of the polarization.],

f. Diameter of the point source (i.e. single bead),

g. Laser distribution in the objective pupil (normally uniform or Gaussian),

h. Correlation coefficient for the goodness of fit for the Gaussian curve.

## METHOD FOR MEASURING STRENGTH OF OPTICAL SECTIONING (SECTION 4.1.3)

From the confocal ISO, this is determined by the FWHM of the signal from a thin uniform fluorescent layer OR a reflective planar object scanned through the focus.

In this method, we propose using a reflective planar object (a mirrored glass slide), as this is more easily acquired, cheaper and more stable than a thin fluorescent layer (e.g. Brakenhoff slide). These can either be prepared using a simple Tollens reaction, or sputter coating in an EM unit, or purchased quite cheaply and mounted on a slide (e.g. https://www.edmundoptics.com/f/first-surface-mirrors/12017/).

1. Place your mirrored slide on the stage and focus on the surface. This can most easily be performed using a low power objective with the transmission lamp and finding small holes in the mirrored surface or finding the edge of the mirrored area. Then move to the high-powered objective requiring QC and adjust focus, again using holes in the surface if possible. If no holes are obvious (new slides are often very well coated), then adjust in the step below.

2. Set up the acquisition parameters so that the reflected laser is captured by one of the detectors. Take care to ensure that the laser power and detector Gain are kept low. Fine tune the focal plane until the brightest point is reached- this is the outside of the mirror surface. Ensure there is no saturation but that you are using the full dynamic range available. Create a z stack, sampling 10-fold more frequently than the theoretical axial resolution (400 nm for a 1.4 NA oil lens at 520 nm emission, so every 40 nm), covering 10 µm.

3. For analysis (section 4.1.4), a Gaussian fit should be made for a single pixel width line profile axially through the stack from the centre of the field. Use of the ImageJ plugin, PSFj (or the standalone version: http://www.knoplab.de/psfj/download/) is recommended as it will compare the axial resolution across the entire field from these images (http://www.nature.com/nmeth/journal/v11/n10/full/nmeth.3102.html). Alternatively, analysis using the ImageJ/ Fiji plugin MetroloJ and using the axial resolution report function. Further details and the manual are available at: https://imagejdocu.tudor.lu/plugin/analysis/metroloj/start.

4. To satisfy the confocal ISO, confocal FWHM should be reported alongside other metadata:

    a. excitation wavelength,

    b. detection wavelength band,

    c. objective designation from manufacturer,

    d. pinhole size in Airy units (plus the reference wavelength used to calculate its size),

    e. Polarisation state of excitation light plus its direction relative to the axis of the determined lateral FWHM,

  f. Diameter of the point source (ie bead),

  g. Laser distribution in the objective pupil (normally uniform or Gaussian),

  h. Correlation coefficient for the goodness of fit for the Gaussian curve.

  i. The type of planar object -reflective or fluorescent

## ADDITIONAL MEASUREMENTS

Although not specified in the confocal ISO, we feel that the additional measurements below would complement it and provide a quick and robust method to monitor system performance over time.

### LASER/ LIGHT POWER AT OBJECTIVE (10X)

This allows monitoring of the light source combined with the whole excitation lightpath, and will allow comparison of values over time. This will highlight any drop off in laser/lamp excitation at the specimen, which identifies either light source fault or problems within the lightpath (eg damaged filters/ dichroics).

**Requirements: calibrated power meter and sensor covering the wavelengths and power of the system.**

#### METHOD

A suitable power meter (as described above) is placed in the focal plane of the standard 10x objective on the system (usually air, 0.3-0.4 NA). Power for each light source is recorded at a range of intensities with a fixed set of dichroics either with spot scanning or a zoomed small area scan (e.g. 5x, 256 x 256). For widefields, this is done on each filter cube with a range of lamp intensities. Minimum standard of 10 and 100% light source power.

Alternative method 1: the objective could be removed, and one would measure the light coming out the same way as if an objective were present. In this way, the measure is independent of the presence of the right objective and cleanness of the objective. There would also be less risk of damaging the surface of the power sensor.

Alternative method 2: using Argolight Power slide using 10x objective recorded at 3 powers (1, 50 and 100% laser power) with a minimum of 5 recordings at each power. If spot scanning isn't possible, ensure all other settings are maintained constant (e.g. 5x zoom, 256x256 bidirectional, max speed).

### LASER/ LIGHT POWER AT SOURCE

This allows monitoring of the light source alone, allowing identification of dying lasers/ lamps, irrespective of any possible confounding factors from the downstream light path.

**Requirements: calibrated power meter mounted in the beam path (ideally via access to laser bed).**

Ideally, this would be performed with an *in- situ* take-off, e.g. via a 99/1 mirror to a permanently housed sensor, or the manufacturer of the laser combiner would prepare a slit where a power sensor could be inserted. This would avoid plugging out and in the laser fibre in the scan head, which can damage the end of the fibre.

*[Note: can only be done on systems with access to the beam path without moving it e.g. spinning disk laser bed, TIRF laser bed. Need to ensure highest recorded output is within power sensor range. Ideally, the manufacturer would provide this as part of the system with a dedicated power measurement output. This only needs to be a value that can be recorded for the system to track changes, not an actual output from the scan head. An alternative would be to add a sensor in place of an objective- this is less ideal, but a simple way for an end user to get values that are independent of the lens (but still have light passing through the entire system).]*

## METHOD OVERVIEW

A power measurement is made in the beam path before it enters the microscope optics/ scan head. Minimum standard of a single averaged measurement for each light source, but ideally longer-term measurements to identify fluctuations should be performed less frequently too.

## DETECTOR LINEARITY

This allows monitoring of detector sensitivity and linearity, and with temporal sampling would allow early identification of faulty detectors (e.g. degraded PMT, damaged camera vacuum chamber). These data, combined with excitation power at the objective, would allow accurate repetition of image acquisition parameters, improving data reproducibility within and between laboratories.

## METHOD OVERVIEW

Each detector/ camera is recorded at fixed settings using a range of excitation intensities. Ideally, this is performed with a standardised external light source mounted at the objective but could also be performed using the internal light sources and a mirror slide. A minimum standard would be recording down to lowest detection sensitivity and up to saturation with at least three measurements between. Ideally, this would be performed across a range of detector gains and using a calibrated external light source.

Alternative method 1. one would do these measurements with fixed excitation powers set with a power sensor placed at the objective. No need for a standardised external light source, as long as the power meter is properly calibrated. Using the Argolight slide would make it easier than using a mirror, as one gets 16 levels at once, but this solution is linked to a commercial product. Using Argolight Power slide, the power sensor is used to determine the excitation power at the objective and modulated to a pre-set value (see power at objective section above). Then using the 16 grey level pattern, intensity range is recorded at fixed exposure/gains to cover minimal detection to saturation of the detectors (may require 2 or 3 different values, and these should be re-used each time once set).

## DETECTOR SENSITIVITY WITH WAVELENGTH

This allows monitoring of detector sensitivity over the spectrum. In combination with detector linearity and excitation power at the objective, this would allow accurate repetition of image acquisition parameters between microscope systems, improving data reproducibility within and between laboratories.

Do we expect the sensitivity to change over time differently for the different wavelengths? if you do not measure it you don't know! we would not expect this to change too much, but it would be good to check periodically.

*[Note: We've enquired with PMT manufacturers (Hamamatsu) as regards sensitivity over time measurements, they have not made them but do not expect their spectral sensitivity to change over time. We're not too sure if this transposes to HyDs or GaAsPs.]*

### METHOD OVERVIEW

Each detector/ camera is recorded at fixed settings using a range of excitation wavelengths. Ideally, this is performed with a standardised external light source mounted at the objective but could also be performed using the internal light sources and a mirror slide. A minimum standard would be recording the longest and shortest wavelengths available on the system. Ideally, this would be performed across a range of wavelengths and using a calibrated external light source.

## Z DRIVE ACCURACY

Accurate, reproducible movement of the stage/objective nosepiece is essential when rendering data in 3D and being able to accurately ascertain where objects labelled at different wavelengths are relative to one another and/or movement in time through z.

### METHOD OVERVIEW

Determine the accuracy of the z drive movement by moving with small steps through a defined 3D landscape. Z drive repositioning can also be tested by performing a short time series of 3D stacks and checking for drift or repositioning error. A defined 3D landscape that can be repeatedly looked at over months/years is required- an etched pattern is the best option in this case, e.g. Argolight or PSFCheck slide. Using the Argolight Power slide, the 3D crossing stairs pattern can be measured at one wavelength, at Nyquist z sampling and captured as 5 timepoint timeseries. Accuracy of positioning and repositioning can be determined from the resulting data.

## STAGE REPOSITIONING ACCURACY

Accurate, reproducible movement of the stage is essential for tile stitching and measuring movement of a sample when capturing multiple positions over time.

### METHOD OVERVIEW

Determine the stage repositioning accuracy of the x-y stage movement by moving between multiple fixed points over time. The distance moved should be large enough to cover the sort of distances expected of the stage movement in experiments. Ideally this would be the size of a multi-well plate for many automated systems. Using the Argolight Power slide or the PSFCheck slide, 3 or more of the positioning crosses can be imaged, ensuring the stage moves in both x and y directions. By running a small time series to capture multiple positions over time the lateral repositioning drift over time can be determined in both dimensions.

# APPENDIX

## 1. SAMPLE PREPARATION

Taken and adapted from PSFj manual: http://www.knoplab.de/psfj/online-manual/

### GLASS SLIDE AND COVERSLIPS

Cover slip thickness needs to adhere tightly to the value specified for the objective (usually 170 μm thick– #1.5 or 1.5H for a pre-selected narrower range of thicknesses). In addition, beads should be attached to the coverslip and not to the glass slide since the latter will position the beads a few micrometers away from the cover slip glass surface when imaged through the cover slip (notable exception: water dipping lenses).

### FLUORESCENT BEADS

For direct extraction of PSFs from an image stack of fluorescent beads (use multicolour beads if you intend to measure chromatic aberrations), the beads should ideally be as small as possible. However, the smaller the beads the lower the signal-to-noise ratio (for a given exposure time) and the more difficult the quantification becomes. However, for resolution measurements, 100nm beads are recommended, and using Tetraspeck multicolour beads allows one to use the same sample for co-registration and lateral PSF measurements.
(https://www.thermofisher.com/order/catalog/product/T7279#/T7279).  If fluorescence intensity is too weak, then slightly larger beads can be used for co-registration analysis.  Also, check that your beads are essentially aggregate free. If necessary, try to disperse these aggregates using a water bath sonicator. The beads should also exhibit a narrow size distribution, i.e. a small coefficient of variation of their diameter (check the specification of the manufacturer). For preparation of the slide, follow the manufacturer's protocol, which usually includes the following steps:

1. Dilute the original bead suspension to a suitable concentration (around $10^9$ beads per ml). Bead density is critical: if the density is too high, this will cause that many beads will be discarded from the analysis. If the density is too low, only a few beads are available for analysis and the statistics will be poor. Manually prepared bead slides show heterogeneous distributions of the beads, so that image stacks with higher or lower bead densities can be recorded and analysed.

2. Sonicate the bead solution (using a water bath sonicator) in order to dissociate possible aggregates.

3. Wash a #1.5 (170 µm thick) coverslip using 70% ethanol (please note: some objective lenses require coverslips of different thickness. In the case of water-dipping objectives, no coverslip is required; in this case the beads are adsorbed directly on the surface of the glass slide, as described in the next step.

4. Place a small amount (around 5-10 µl) of the bead solution on the coverslip and let it dry.

5. Place a small amount (depending on your coverslip size, around 10 µl for 24×24 mm coverslips) of mounting medium on the coverslip and place the slide upside down onto a microscope slide.

6. Seal the gap between coverslip and microscope slide, e.g. using non-fluorescent nail polish.

*[NOTE:  Dust/dirt particles, but also too much or too little mounting medium can produce a coverslip that is not aligned parallel to the surface of the glass slide. If this is the case, the*

*coverslip surface with the beads may not be orthogonal to the optical axis of the microscope. Consequently, a linear gradient (along the direction of the tilt) will appear in the heat map that visualizes the planarity of the system.]*

## 2. THEORETICAL RESOLUTION

Example resolutions for common high-end oil immersion objectives are given in the ISO document, examples for oil immersion objectives are given below, with the assumption of 488nm excitation and 520 nm emission with a pinhole set at 1 Airy unit. Resolutions reported in µm:

Oil immersion (assuming refractive index 1.518)

| NA | Lateral Resolution | Axial Resolution | Optical Sectioning |
|---|---|---|---|
| 1.45 | 0.16 | 0.34 | 0.43 |
| 1.40 | 0.16 | 0.40 | 0.49 |
| 1.30 | 0.18 | 0.48 | 0.60 |